\documentclass[twocolumn,showpacs,preprintnumbers,amsmath,amssymb,floatfix]{revtex4}
\usepackage{epsf,epsfig,graphicx}

\newcounter{muni}

\begin{document}
\hbadness=10000 \pagenumbering{arabic}

\title{Fourth Generation CP Violation Effect on $B\to K\pi$, $\phi K$ and $\rho K$ in NLO PQCD}

\author{Wei-Shu Hou$^{1}$}
\author{Hsiang-nan Li$^{2,3}$}
\author{Satoshi Mishima$^4$}
\author{Makiko Nagashima$^5$}

 \affiliation{$^{1}$Department of Physics, National Taiwan
University, Taipei, Taiwan 106, Republic of China}
 \affiliation{$^{2}$Institute of Physics, Academia Sinica, Taipei,
Taiwan 115, Republic of China}
 \affiliation{$^{3}$Department of
Physics, National Cheng-Kung University, Tainan, Taiwan 701,
Republic of China}
 \affiliation{$^{4}$School of Natural Sciences,
Institute for Advanced Study, Princeton, NJ 08540, U.S.A.}
 \affiliation{$^5$Physique des Particules, Universit\'e
de Montr\'eal, 
Montr\'eal, 
Quebec, Canada H3C 3J7}

\begin{abstract}

We study the effect from a sequential fourth generation quark on
penguin-dominated two-body nonleptonic $B$ meson decays in the
next-to-leading order perturbative QCD formalism. With an
enhancement of the color-suppressed tree amplitude and possibility
of a new CP phase in the electroweak penguin, we can account
better for $A_{\rm CP}(B^0\to K^+\pi^-)-A_{\rm CP}(B^+\to
K^+\pi^0)$. Taking $|V_{t's}V_{t'b}|\sim 0.02$ with phase just
below $90^\circ$, which are consistent with the $b\to
s\ell^+\ell^-$ rate and the $B_s$ mixing parameter $\Delta
m_{B_s}$, we find a downward shift in the mixing-induced CP
asymmetries of $B^0\to K_S \pi^0$ and $\phi K_S$. The predicted
behavior for $B^0\to\rho^0 K_S$ is opposite.
\end{abstract}


\maketitle

CP violation (CPV) in $b\to s$ transitions is at the forefront of
our quest to understand
flavor and the origins of CPV, offering one of the best probes for
New Physics (NP) beyond the Standard Model (SM). Several hints for
NP have emerged in the past few years.
For example, a large difference is seen in direct CP asymmetries
in $B\to K\pi$ decays~\cite{HFAG},
\begin{eqnarray}
{\cal A}_{K\pi}
 \equiv A_{\rm CP}(B^0\to K^+\pi^-) = -0.093 \pm 0.015, &&\nonumber\\
{\cal A}_{K\pi^0}
 \equiv A_{\rm CP}(B^+\to K^+\pi^0) = +0.047 \pm 0.026, &&
 \label{data}
\end{eqnarray}
or $\Delta{\cal A}_{K\pi} \equiv {\cal A}_{K\pi^0}-{\cal A}_{K\pi}
= (14\pm 3)\%$~\cite{Barlow}. As it was not predicted when first
measured in 2004, it has stimulated
discussion on the potential mechanisms that may have been missed
in the SM calculations \cite{BBNS,KLS,BPS05}.

Better known is the mixing-induced CP asymmetry ${\cal S}_f$
measured in a multitude of CP eigenstates $f$.
For penguin-dominated $b \to sq\bar q$ modes, within SM, ${\cal
S}_{sq\bar q}$ should be close to that extracted from $b\to c\bar
cs$ modes. The latter is now measured rather precisely, ${\cal
S}_{c\bar cs}=\sin2\phi_1 = 0.674 \pm 0.026$~\cite{Hazumi}, where
$\phi_1$ is the weak phase in $V_{td}$. However, for the past few
years, data seem to indicate, at 2.6$\,\sigma$ significance,
\begin{eqnarray}
\Delta {\cal S} \equiv {\cal S}_{sq\bar q}-{\cal S}_{c\bar cs}
\lesssim 0,
 \label{DelS}
\end{eqnarray}
which has stimulated even more discussions.

Since the two modes in Eq. (\ref{data}) differ by the subleading
color-suppressed tree amplitude $C'$ and electroweak penguin
amplitude $P_{\rm EW}'$, it is natural to conjecture that $C'$,
$P_{\rm EW}'$, or maybe both have been underestimated (see
\cite{Charng2} and references therein). In this paper we study the
fourth generation effect on the $B\to K\pi$, $\phi K$ and $\rho K$
decays in the perturbative QCD (PQCD) approach, which
exploits both
$C'$ and $P_{\rm EW}'$ to account for Eq.~(\ref{data}), and gives
the right tendency towards explaining Eq.~(\ref{DelS}).

At leading order (LO), PQCD predicted~\cite{KLS} the sign and
strength of ${\cal A}_{K\pi}$, but ${\cal A}_{K\pi^0}$ was also
negative and not too different from ${\cal A}_{K\pi}$.
Going to next-to-leading-order (NLO)~\cite{LMS05}, it was shown
that inclusion of the vertex corrections increases $C'$ by a
factor of 3, without affecting much the branching ratios, ${\cal
A}_{K\pi^0}$ becomes much closer to zero, although still negative.
However, like many other studies in the literature
\cite{CGRS,CCS2,B05}, NLO PQCD also predicts $\Delta {\cal
S}_{K_S\pi^0} \equiv {\cal S}_{K_S\pi^0} - {\cal S}_{c\bar cs} >
0$ within SM, opposite to what is indicated by Eq.~(\ref{DelS}).
Thus, if $\Delta{\cal S}$ stands the scrutiny of time, one would
still need to go beyond SM.

The other path to account for Eq. (\ref{data}), to have an effect
coming from $P_{\rm EW}'$, in fact requires NP CPV. Among the
available models, the sequential fourth generation is found to
modify only $P_{\rm EW}'$ significantly~\cite{HNS}, but not other
amplitudes such as the QCD penguin $P'$. With reasonable
parameters, e.g. with $m_{t'} \simeq 300$ GeV, the product of
quark mixing matrix elements $|V_{t's}V_{t'b}| = 0.01\sim 0.03$
with phase close to $90^\circ$, it was found in PQCD at LO that
${\cal A}_{K\pi^0}$ could also be rendered vanishing. Remarkably,
the constraints of the $b\to s\ell^+\ell^-$ rate and the $B_s$
mixing parameter $\Delta m_{B_s}$ are also satisfied.
It was shown~\cite{HNRS} in QCD factorization (QCDF) at NLO (and
PQCD at LO for $B\to K \pi$), that the same parameter choice
produces a downward shift for ${\cal S}_{K_S\pi^0}$ and ${\cal
S}_{\phi K_S}$. The prediction is far from reliable, however, due
to uncontrollable hadronic uncertainties in QCDF.

It is worthwhile, then, to reanalyze the fourth generation effect
using the NLO PQCD formalism, in part to find whether the
preferred parameter space~\cite{HNS} is affected, and also to
investigate the efficacy of combining the $C'$ and $P_{\rm EW}'$
approaches on CPV in penguin-dominated $b\to s\bar qq$ modes. It
should be emphasized that both the strength and phase of
$V_{t's}^\ast V_{t'b}$ would be effectively pinned down by the
recent precise measurement of $\Delta m_{B_s}$ \cite{dmsCDF},
which will then lead to rather definite predictions for CPV in
$B_s$ mixing, $\sin 2\Phi_{B_s}$, as well as $D^0$ mixing
\cite{HNSBsDmix}. We show that, using the same fourth generation
parameters as before, we can explain the trend of the observed
$\Delta S$ in $B\to K\pi$ and $\phi K$ decays simultaneously,
while understanding of $\Delta {\cal A}_{K\pi}$ is also improved.
We point out that the fourth generation effect is opposite in
$B\to\rho K$, which increases $S_{\rho^0 K_S}$ from a low value in
SM \cite{B05,LM06}.

It is instructive to elucidate the underpinnings of the two
proposals to resolve the $B\to K\pi$ puzzle, i.e. enhanced $C'$ by
vertex corrections in SM, vs fourth generation with a new CPV
phase.

The NLO PQCD calculation within SM shows~\cite{LMS05} that $P'$ is
in the second quadrant, and the color-allowed tree amplitude $T'$
is roughly real and positive. Enhanced by the vertex corrections,
$C'$ turns almost imaginary, and $T'+C'$ is in the fourth quadrant
and almost opposite in direction as $P'+P_{\rm EW}'$. As ${\cal
A}_{K\pi^0}$ is proportional to the sine of the angle between
$T'+C'$ and $P'+P_{\rm EW}'$, its value drops in NLO PQCD within
SM.
In the four generation model at LO in PQCD~\cite{HNS}, the
amplitude $P_{\rm EW}'$ acquires a large CPV phase from
$V_{t's}^*V_{t'b}$, which effectively cancels the weak phase
$\phi_3\equiv \arg V_{ub}^\ast$ in $T'+C'$, making ${\cal
A}_{K\pi^0}$ almost vanish. The $B^0 \to K^+ \pi^-$ decay,
however, is less affected, as the $P_{\rm EW}'$ contribution is
color-suppressed. 
Since only subleading amplitudes are modified, in both proposals
the $B\to K\pi$ branching ratios are not much affected. This is
contrary to the proposals based on either inelastic \cite{CCS} or
elastic \cite{CHY} final state rescattering, in which leading $P'$
amplitudes are enhanced. This is the reason why the $B\to K\pi$
puzzle cannot be resolved in such approaches.

For mixing-induced CP asymmetries, there is clearly a major
difference between the above proposals. The time-dependent CP
asymmetry of the $B^0\to K_S\pi^0$ mode is defined as
\begin{eqnarray}
&& A_{\rm CP}(B^0(t)\to K_S\pi^0)
\nonumber\\
 &\equiv& \frac{{\cal B}({\bar B}^0(t)\to
K_S\pi^0)- {\cal B}(B^0(t)\to K_S\pi^0)} {{\cal B}({\bar
B}^0(t)\to K_S\pi^0)+{\cal B}(B^0(t)\to K_S\pi^0)}
\nonumber\\
&=& {\cal A}_{K_S\pi^0}\cos\Delta m_{B_d} t + {\cal
S}_{K_S\pi^0}\sin\Delta m_{B_d} t,
 \label{CPk}
\end{eqnarray}
where $\Delta m_{B_d}$ is the mass difference of the two $B$-meson
mass eigenstates, and
\begin{eqnarray}
{\cal A}_{K_S\pi^0} = {|\lambda_{K_S\pi^0}|^2-1 \over
1+|\lambda_{K_S\pi^0}|^2},
\,\ {\cal S}_{K_S\pi^0} = {2\,{\rm Im}\,\lambda_{K_S\pi^0} \over
1+|\lambda_{K_S\pi^0}|^2},
 \label{spk}
\end{eqnarray}
are the direct and mixing-induced asymmetries, respectively. The
$B^0\to K_S\pi^0$ decay has a CP-odd final state, and the
associated factor,
\begin{eqnarray}
\lambda_{K_S\pi^0}\, =\, -e^{-2i\phi_1}
 {{\cal M}({\bar B}^0\to K_S\pi^0) \over
  {\cal M}({     B}^0\to K_S\pi^0)},
 \label{mix}
\end{eqnarray}
with ${\cal M}({\bar B}^0\to K_S\pi^0) = P' -P'_{\rm EW}
-C'e^{-i\phi_3}$, where we have isolated the weak phase in
SM~\cite{foot}.

Although $C'$ can be enhanced by a few times from the vertex
corrections in NLO PQCD, within SM
${\cal S}_{K_S\pi^0}$ is not much affected, which stays close to
${\cal S}_{c\bar cs}$. According to Eq.~(\ref{mix}), the leading
deviation caused by $C'$ is proportional to the cosine of the
relative strong phase between $C'$ and $P'-P_{\rm EW}'$. Because
the vertex corrections also rotate $C'$, it becomes more
orthogonal to $P'$, and the cosine diminishes.
However, in the fourth generation model, a NP phase is introduced
through $P_{\rm EW}'$, so that $\Delta {\cal S}_{K_S\pi^0}$ could
be sizable. As we will show, the simultaneous accommodation of
$\Delta{\cal A}_{K\pi}$ and $\Delta {\cal S}_{K_S\pi^0}$ is
nontrivial, with all experimental constraints suitably satisfied.

\begin{figure*}[t!]
\smallskip  
\includegraphics[width=4.1cm,height=2.85cm,angle=0]{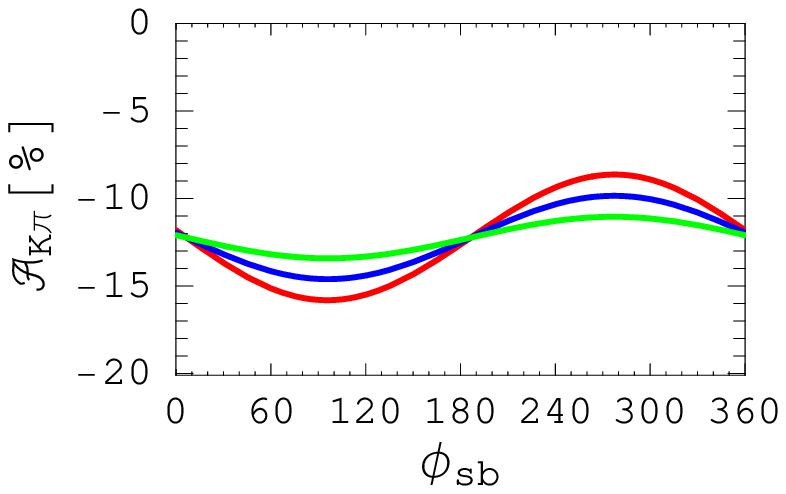}
\includegraphics[width=4.1cm,height=2.85cm,angle=0]{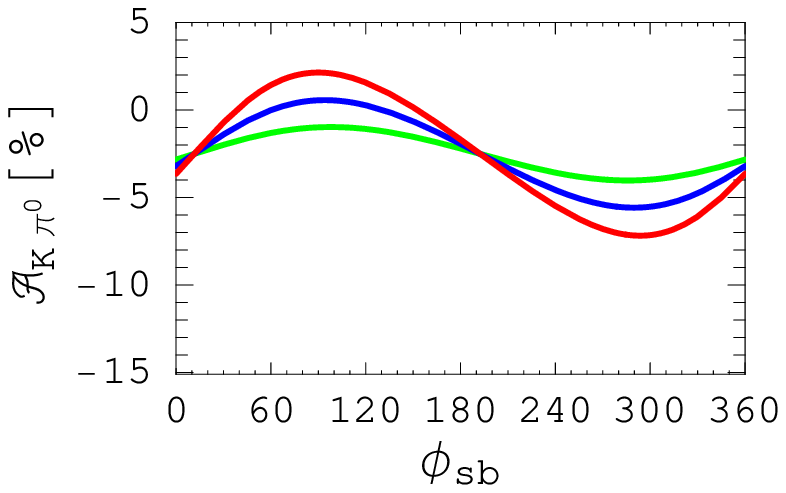}
\includegraphics[width=4cm,height=2.8cm,angle=0]{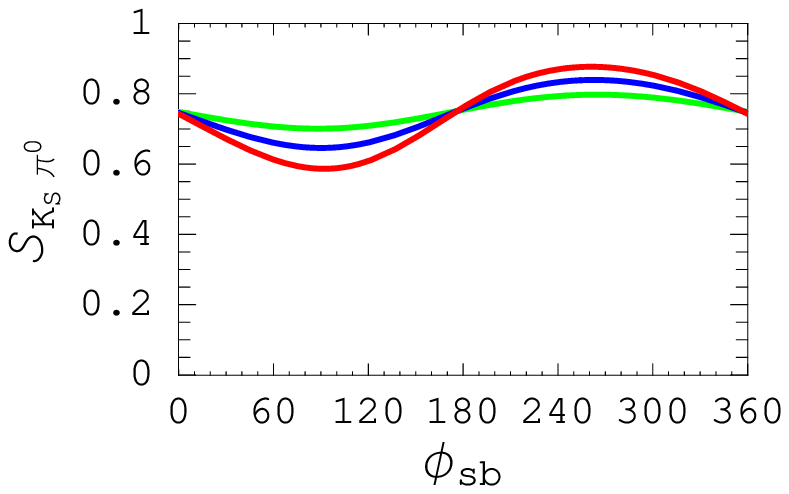}
\includegraphics[width=4cm,height=2.8cm,angle=0]{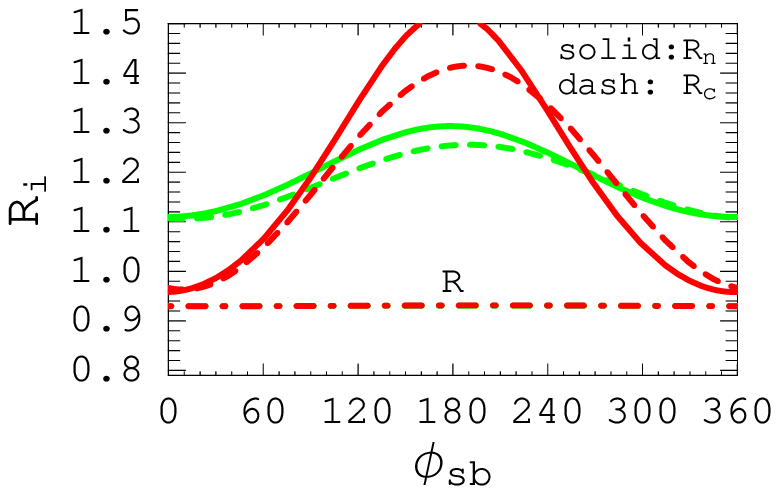}
 \caption{
  (a) ${\cal A}_{K\pi}$,
  (b) ${\cal A}_{K\pi^0}$,
  (c) ${\cal S}_{K_S\pi^0}$ and
  (d) $R_i$ ($i={\rm null},\,c,\,n$) vs $\phi_{sb}$ for $m_{t^\prime} = 300$ GeV
  with all of the NLO corrections.
  The curves are for $r_{sb}= 0.01$, 0.02 (not shown in (d)), 0.03,
  with $r_{sb}=0.03$ giving the strongest effect.
 \label{fig:Fig1}
}
\end{figure*}

The NLO PQCD formalism for the $B\to PP$, $PV$ decays, in which
the contributions from the NLO evolution of Wilson coefficients,
the vertex corrections, the quark loops, and the magnetic penguin
are taken into account, can be found in Refs.~\cite{LMS05} and
\cite{LM06}. The main assumption involved is that we have
neglected NLO corrections to $B$ meson transition form factors,
which is not essential for studies of CP asymmetries. The
procedure to incorporate the fourth generation effect can be found
in Ref.~\cite{HNS}.
In Figs.~\ref{fig:Fig1}(a), (b) and (c) we plot the predictions
for ${\cal A}_{K\pi}$, ${\cal A}_{K\pi^0}$ and ${\cal
S}_{K_S\pi^0}$ vs $\phi_{sb} \equiv \arg V_{t's}^*V_{t'b}$, for
three values of $r_{sb} \equiv |V_{t's}^*V_{t'b}| =$ 0.01, 0.02
and 0.03. We have taken $\phi_1=21.6^\circ$, $\phi_{3}=70^\circ$,
and $m_{t'} = 300$ GeV. Theoretical uncertainties from the above
Cabbibo-Kobayashi-Maskawa (CKM) matrix elements and from hadronic
parameters, such as 
Gegenbauer coefficients in meson distribution amplitudes, can be
included straightforwardly as in Ref.~\cite{LMS05}. Considering
these uncertainties, the predicted ranges of ${\cal A}$ and ${\cal
S}$ will be enlarged by about 50\% and a few percent,
respectively.
A variation in $m_{t'}$ just changes the range of $r_{sb}$ (i.e.
the CKM product $|V_{t's}^*V_{t'b}|$). Indeed, ${\cal A}_{K\pi^0}$
becomes more positive for $\phi_{sb} \approx 90^\circ$, while
${\cal S}_{K_S\pi^0}$ dips downwards, showing that the nearly
vanishing ${\cal A}_{K\pi^0}$ and the smaller ${\cal
S}_{K_S\pi^0}$ occur at the same NP phase. Compared with
Ref.~\cite{HNS} in LO PQCD, we observe that the strength of the
parameter $r_{sb}$ has shrunk: ${\cal A}_{K\pi^0}$ flips sign for
$m_{t'}=300$ GeV and $r_{sb}=0.03$ in \cite{HNS}, but now for
$r_{sb}=0.02$. This implies that the larger color-suppressed tree
amplitude in NLO PQCD has softened the need for larger NP
parameters. The decrease of $R_n\approx 1.1$ for $\phi_{sb}
\approx 90^\circ$ shown in Fig.~\ref{fig:Fig1}(d) is now
consistent with the observed value.

\begin{table}[b!]
\begin{center}
\begin{tabular}{crrrrr}
\hline\hline Quantity & +WC & +VC & +QL &  +MP  & +NLO (SM)
\\
\hline 
${\cal B}(K^0\pi^\pm)$ &
 $23.6$&$22.8$&$25.2$&$17.7$&
 $18.3$ ($18.1$)
\\
${\cal B}(K^\pm\pi^0)$ &
 $12.6$&$12.2$&$13.4$&$9.6$&
 $9.9$ ($10.7$)
\\
${\cal B}(K^\pm\pi^\mp)$ &
 $20.3$&$19.8$&$21.6$&$15.2$&
 $15.8$ ($15.7$)
\\
${\cal B}(K^0\pi^0)$ &
 $9.3$&$9.1$&$10.0$&$\phantom{-}6.8$&
 $7.2\,$ \ ($6.5$)
\\
\hline \hline  
${\cal A}_{K^0\pi}$ &
 $0$&$1$& $0$& $0$&
 $2\phantom{-}\,$ \ ($0$)
\\
${\cal A}_{K\pi^0}$ &
 $-5$&$1$&$-4$&$-7$&
 $1\,$ \ ($-2$)
\\
${\cal A}_{K\pi}$ &
 $-11$&$-13$&$-9$&$-14$&
 $-15$ ($-12$)
\\
${\cal A}_{K^0\pi^0}$ &
 $-5$& $-12$&$-4$& $-7$&
 $-14\,$ \ ($-8$)
\\
\hline\hline
\end{tabular}
\end{center}
\caption{
  Branching ratios (in unit of $10^{-6}$) and direct CP asymmetries
  (in unit of $10^{-2}$) for $B \to K\pi$ decays,
  with $m_{t^\prime} = 300$ GeV, $r_{sb}=0.025$ and
  $\phi_{sb}=65^\circ$~\cite{foot1}.
  The labels +WC, +VC, +QL, +MP, and +NLO mean the LO results with the NLO
  Wilson coefficients, the
  inclusions of the vertex corrections, of the quark loops, of the
  magnetic penguin, and of all the above NLO corrections,
  respectively. The NLO predictions from SM are presented in the
  parentheses for comparison. }\label{RateDCPVpiK}
\end{table}

The rates and ${\cal A}_{\rm CP}$s of all four $B\to K\pi$ modes
are listed in Table~\ref{RateDCPVpiK}, which elucidates the
effects from the NLO Wilson evolution, vertex corrections, quark
loops, and the magnetic penguin, separately. The tendency is
basically the same as in Ref.~\cite{LMS05}.
The main change from the incorporation of the fourth generation
appears in the direct CP asymmetries: ${\cal A}_{K\pi^0}$ moves
positive (flips sign) but remains small, and ${\cal
A}_{K^0\pi^0}$, which also depends on $P_{\rm EW}'$, becomes
roughly equal to ${\cal A}_{K\pi}$.
Note that ${\cal A}_{K\pi}$ and ${\cal A}_{K\pi^0}$ are not in
perfect match with Eq.~(1), but varying hadronic parameters such
as meson distribution amplitudes, the rates, ${\cal A}_{K\pi}$,
and ${\cal A}_{K\pi^0}$ could approach data. We do not attempt any
such tuning at present, but took the input parameters from earlier
NLO PQCD analyses that did not involve the fourth generation.

\begin{figure}[b!]
\smallskip  
\includegraphics[width=4cm,height=2.8cm,angle=0]{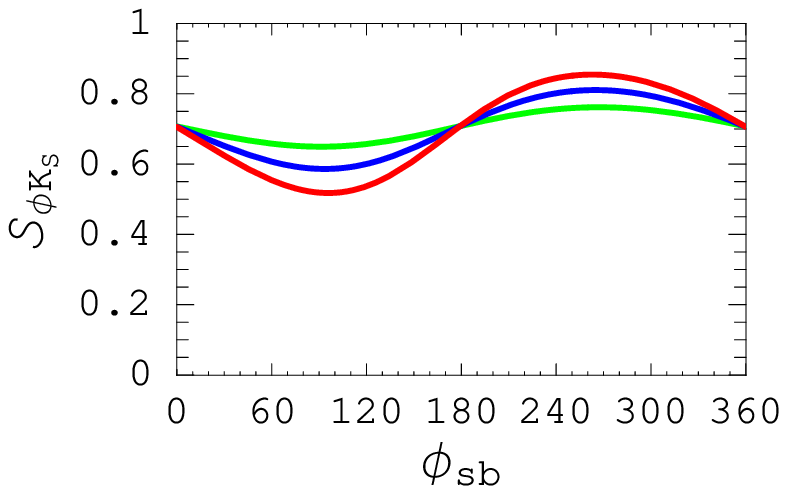}
\includegraphics[width=4cm,height=2.8cm,angle=0]{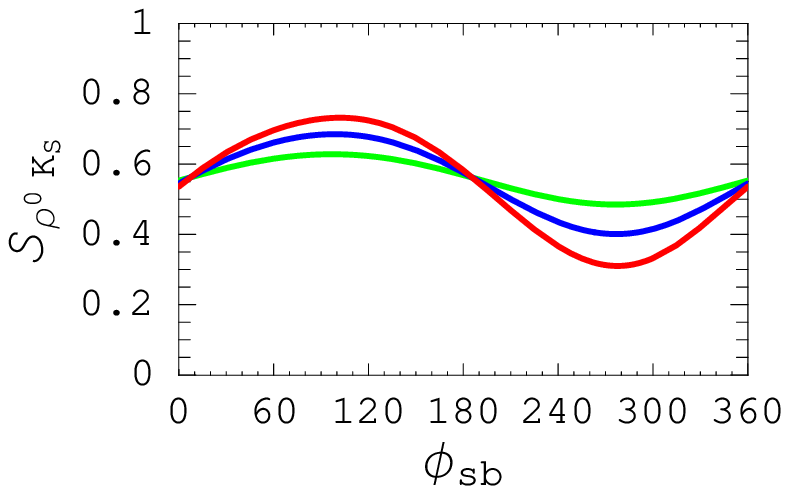}
 \caption{
   (a) ${\cal S}_{\phi K_S}$ and (b) ${\cal S}_{\rho^0 K_S}$ vs
   $\phi_{sb}$ with all of the NLO corrections.
   Notation is the same as Fig.~\ref{fig:Fig1}.
 \label{fig:Fig2}
}
\end{figure}

The predictions for the mixing-induced CP asymmetries are
relatively insensitive to the hadronic parameters. The results for
${\cal S}_{\phi K_S}$ and ${\cal S}_{\rho^0 K_S}$ are displayed in
Fig.~\ref{fig:Fig2}. The effect for ${\cal S}_{\phi K_S}$ is
similar to ${\cal S}_{K_S\pi^0}$, going down from ${\cal S}_{\phi
K_S} \simeq \sin2\phi_1$ in SM. However, ${\cal S}_{\rho^0 K_S}$
behaves in an opposite way: it gets enhanced from 0.5 in SM
\cite{LM06}, with Eq.~(\ref{DelS}) barely obeyed. The rates and
${\cal A}_{\rm CP}$s of the $B\to\phi K$ and $\rho K$ decays are
collected in Table~\ref{RatephiK}. Compared to the NLO results
from SM, the branching ratios are not affected much by the fourth
generation as stated before. The change mainly appears in the
direct CP asymmetries:
 ${\cal A}_{\phi K}$ and ${\cal A}_{\phi K^0}$, though small, turn
 negative;
 ${\cal A}_{\rho^0K}$ decreases like ${\cal A}_{K\pi^0}$,
 but is still sizable;
 ${\cal A}_{\rho^0 K^0}$ increases like ${\cal A}_{K^0\pi^0}$, and
 becomes comparable with ${\cal A}_{\rho^0K}$ and ${\cal A}_{\rho K}$.

We give a general explanation on the above findings, assuming that
NP at high energy modifies the Wilson coefficients relevant to the
electroweak penguin in the low-energy effective theory (SM). Both
the four generation model and the $Z'$ model \cite{BC} provide
such examples. The NP effect on the direct CP asymmetries and
mixing-induced CP asymmetries is written, up to corrections of
$O(r_{\rm EW}^2)$, as
\begin{eqnarray}
\Delta {\cal A}_{K\pi^0}
 &=&  2r_{\rm EW}\sin\delta_{\rm EW}\sin\phi_{\rm EW},\nonumber\\
\Delta {\cal S}_{K_S\pi^0}
 &=& -2r_{\rm EW}\cos2\phi_1 \cos\delta_{\rm EW} \sin\phi_{\rm EW},
\end{eqnarray}
where $r_{\rm EW}$ is the ratio of the magnitude of the NP
contribution to the electroweak penguin over the full penguin
amplitude, $\delta_{\rm EW}$ the relative strong phase between the
electroweak penguin and the full penguin in SM, and $\phi_{\rm
EW}$ the NP phase. Since $P'_{\rm EW}$ is anti-parallel to $T'+C'$
from isospin symmetry \cite{NGW}, 
$\delta_{\rm EW}$ is expected to be less than $90^\circ$. Hence,
the NP phase $\phi_{\rm EW}<90^\circ$ leads to a decrease of the
magnitude of both ${\cal A}_{K\pi^0}$ (if ${\cal A}_{K\pi^0}<0$ in
SM, such as in the PQCD approach) and ${\cal S}_{K_S\pi^0}$. That
is, the current data of ${\cal S}_{K_S\pi^0}$ favor the presence
of NP in the electroweak penguin. Adopting the same parameters, NP
increases ${\cal S}_{\rho^0 K_S}$ as shown in Fig.~\ref{fig:Fig2},
since the NLO PQCD analysis has indicated $\delta_{\rm EW} >
90^\circ$ in the $B\to\rho K$ decays \cite{LM06}, which is
attributed to the change of the penguin amplitude with destructive
combination of the Wilson coefficients $a_4$ and $a_6$. If the
pattern of the direct CP asymmetries ${\cal A}_{\rho K}\approx
{\cal A}_{\rho^0 K}$ \cite{LM06}, related to the orientation of
the penguin amplitude, is confirmed, the increase of ${\cal
S}_{\rho^0 K_S}$ will become solid.  Note that implementing NP
into other theoretical approaches may not resolve the $\Delta{\cal
A}_{K\pi}$ and $\Delta{\cal S}_{K_S\pi^0}$ puzzles, and may not
lead to the increase of ${\cal S}_{\rho^0 K_S}$.
The recent first measurement of ${\cal S}_{\rho^0
K_S}$~\cite{HFAG,Hazumi} gives a low value, but with rather large
errors.

\begin{table}[t!]
\begin{center}
\begin{tabular}{crrrrr}
\hline\hline Quantity & +WC & +VC & +QL &  +MP  & +NLO (SM)
\\
\hline 
${\cal B}(\phi K^\pm)$ &
 $31.0$&$12.9$&$33.0$&$21.7$&
 $\phantom{-}8.1$ ($7.2$)
\\
${\cal B}(\phi K^0)$ &
 $28.8$&$12.0$&$30.7$&$20.2$&
 $\phantom{-}7.5$ ($6.8$)
\\
\hline  
 ${\cal B}(\rho^\pm K^0)$ &
 $\phantom{-}5.4$&$\phantom{-}7.3$&$\phantom{-}5.3$&
 $\phantom{-}6.0$&$\phantom{-}7.7$ ($7.7$)
\\
${\cal B}(\rho^0 K^\pm)$ &
 $\phantom{-}2.9$&$\phantom{-}3.7$&$\phantom{-}2.9$&
 $\phantom{-}2.9$&$\phantom{-}3.7$ ($3.9$)
\\
${\cal B}(\rho^\mp K^\pm)$ &
 $\phantom{-}6.0$&$\phantom{-}7.1$&$\phantom{-}5.9$&
 $\phantom{-}7.0$&$\phantom{-}7.8$ ($7.6$)
\\
${\cal B}(\rho^0 K^0) $ &
 $\phantom{-}3.0$&$\phantom{-}3.7$&$\phantom{-}2.9$&
 $\phantom{-}3.7$&$\phantom{-}4.2$ ($4.5$)
\\
\hline\hline ${\cal A}_{\phi K}$ &
 $-3$&$-1$& $-2$& $-5$&
 $-3\phantom{-}$ ($0$)
\\
${\cal A}_{\phi K^0}$ &
 $-3$&$0$&$-2$&$-4$&
 $-2\phantom{-}$ ($2$)
\\
\hline ${\cal A}_{\rho K^0}$ &
 $\phantom{-}4$&$\phantom{-}3$& $\phantom{-}4$& $\phantom{-}5$&
 $4\phantom{-}$ ($1$)
\\
${\cal A}_{\rho^0 K}$ &
 $54$&$45$&$54$&$55$&
 $44\;$ ($67$)
\\
${\cal A}_{\rho K}$ &
 $65$&$62$&$67$&$56$&
 $58$ \ ($57$)
\\
${\cal A}_{\rho^0 K^0}$ &
 $19$& $27$&$0$& $16$&
 $27\,\;$ \ ($6$)
\\
\hline\hline
\end{tabular}
\end{center}
\caption{
  Branching ratios (in unit of $10^{-6}$) and
  direct CP asymmetries (in unit of $10^{-2}$)
  for $B \to \phi K$ and $\rho K$ decays with $m_{t^\prime} = 300$ GeV,
  $r_{sb}=0.025$ and $\phi_{sb}=65^\circ$.
  The notation is the same as Table~\ref{RateDCPVpiK}.
}\label{RatephiK}
\end{table}

In this Letter we have studied the effect of a fourth generation
in the NLO PQCD framework. Combining the enhancement of the
color-suppressed tree with CPV in electroweak penguin renders the
predictions for $\Delta{\cal A}_{K\pi}$ and $\Delta{\cal
S}_{K_S\pi^0}$ more consistent with data, which is a nontrivial
result. We predict several other mixing dependent and direct CPV
effects. Future precise measurements at the B factories, the
Tevatron, the LHC, and (hopefully) Super B factories, will
determine whether NP is called for~\cite{SMH} by $\Delta{\cal S} <
0$, and in turn constrain the NP parameters.


We thank A. Soddu for discussions. This work was supported in part
by Grants No. NSC-95-2112-M-050-MY3, NSC-94-2112-M-002-035,
DE-FG02-90ER40542
and by 
NCTS of the Republic of China.


\begin{thebibliography}{99}
\bibitem{foot}
 Even in the four generation model, there is little deviation from
 $\phi_1$ in $B_d$ mixing, once kaon constraints are taken into
 account~\cite{HNS}.
\end{thebibliography}
\end{document}